\documentstyle[twoside,fleqn,espcrc2,epsf]{article}

\newcommand{\AmS}{{\protect\the\textfont2
  A\kern-.1667em\lower.5ex\hbox{M}\kern-.125emS}}

\title{Results for Quenched $B_K$ from JLQCD\thanks{presented by S. Aoki}}

\author{JLQCD Collaboration : 
	S. Aoki\address{Institute of Physics, University of Tsukuba,
        Tsukuba, Ibaraki 305, Japan},
        M. Fukugita\address{Institute for Cosmic Ray Research, 
        University of Tokyo, Tanashi, Tokyo 188, Japan},
        S. Hashimoto\address{Computing Research Center,
        High Energy Accelerator Research Organization (KEK),\\
        Tsukuba, Ibaraki 305, Japan},
        N. Ishizuka$^{\rm a,}$\address{Center for Computational Physics, 
        University of Tsukuba, Tsukuba, Ibaraki 305, Japan},
	Y. Iwasaki$^{\rm a,d}$,\\
	K. Kanaya$^{\rm a,d}$
	Y. Kuramashi\address{Institute of Particle and Nuclear Studies,
        High Energy Accelerator Research Organization (KEK),
        Tsukuba, Ibaraki 305, Japan},
	M. Okawa$^{\rm e}$,
	A. Ukawa$^{\rm a}$,
	T. Yoshi\'e$^{\rm a,d}$
}
       
\begin{document}

\begin{abstract}
A report is presented on our continued effort to elucidate the continuum
limit of $B_K$ using the quenched Kogut-Susskind quark action.
By adding to our previous simulations one more point 
at $\beta = 6.65$ employing a
$56^3\times 96$ lattice, we now confirm the expected
$O(a^2)$ behavior of $B_K$ with the Kogut-Susskind action.
A simple continuum extrapolation quadratic in $a$ leads to
$B_K$(NDR, 2 GeV) = 0.598(5). As our final value of $B_K$ in
the continuum we present $B_K$(NDR, 2 GeV)=0.628(42), as obtained
by a fit including an $\alpha_{\overline{MS}}(1/a)^2$ term
arising from the lattice-continuum matching with the 
one-loop renormalization.

\end{abstract}

\maketitle

\section{Introduction}

The problem of taking the continuum limit remains to be the most important
issue in the lattice calculation of $B_K$
using the quenched Kogut-Susskind quark action.
We have been making effort to elucidate the effect of scaling violation
over two years since 1995.  Our early study\cite{JLQCD1}
carried out at $\beta = 5.85-6.2$ showed a linear decrease
of $B_K$ in $a$ in contrast to $O(a^2)$ scaling violation
theoretically predicted\cite{sharpe1}.
The run was then extended to $\beta =$ 6.4\cite{JLQCD2},
at which a departure from a linear behavior was observed, 
indicating the onset of $O(a^2)$ behavior. This implied
a cancellation between an $a^2$ and higher order terms that leads to 
an apparent $O(a)$ behavior for lower values of $\beta$. 
We have now extended the run to  $\beta$=6.65 employing a $56^3 \times 96$
lattice to settle the issue of the continuum extrapolation. 
We also briefly address the problem
of $O(\alpha_s^2)$ effect, on which we gained insight after the
present conference.

\section{Perturbative matching}
We have slightly revised the method of analysis in obtaining $B_K$ in
the continuum since Lattice 96\cite{JLQCD2}.
Lattice values of $B_K$ are converted to the continuum value in 
the $\overline{MS}$ scheme with
the naive dimensional regularization (NDR)
by applying one-loop renormalization\cite{IS,SP} at the matching scale 
$q^*=1/a$\cite{Ji,GBS}. 
The one-loop renormalization factor is evaluated with the 3-loop
running coupling constant $\alpha_{\overline{MS}}(q^*)$ 
with $\Lambda_{\overline{MS}}$ = 0.23 GeV, estimated in the continuum
limit from our results for the $\rho$ meson mass.

The continuum value of $B_K$ at the physical scale $\mu =$ 2 GeV is
calculated from $q^*$ via the 2-loop running of the continuum 
renormalization group:
\vspace*{-3mm}
\begin{eqnarray}
& B_K({\rm NDR},\mu)  = \left[
\frac{\displaystyle 1-\displaystyle\frac{\alpha_{\overline{MS}}(q^*)}{4\pi}
\displaystyle\frac{\gamma_1\beta_0-\gamma_0\beta_1}{2\beta_0^2}}
{\displaystyle 1-\displaystyle\frac{\alpha_{\overline{MS}}(\mu)}{4\pi}
\displaystyle\frac{\gamma_1\beta_0-\gamma_0\beta_1}{2\beta_0^2}}
\right]
\nonumber\\
&\times  \left[
{\alpha_{\overline{MS}}(q^*)}/{\alpha_{\overline{MS}}(\mu)}
\right]^{-\gamma_0/2\beta_0} B_K({\rm NDR},q^*)
\nonumber
\end{eqnarray}
with $\beta_0=11$, $\beta_1=102$, $\gamma_0=4$
and $\gamma_1=-7$\cite{BJW}.
This matching procedure leaves  $\alpha_{\overline{MS}}(1/a)^2$ uncertainty,
which however should vanish logarithmically in the continuum 
limit\cite{KGS,Onogi}.

\begin{table}[bt]
\caption{Run parameters and results for $B_K({\rm NDR, 2GeV})$}
\label{tab:run}
\begin{tabular}{lll|ll}
\hline
$\beta$ & $L^3 T$ & $La$
& \multicolumn{2}{c}{$B_K({\rm NDR, 2GeV})$} \\
 & &(fm) & non-inv. & inv. \\
\hline
5.7  & $12^3 24$ &  2.8 & 0.846(1) & 0.822(1)   \\
5.85 & $16^3 32$ &  2.3 & 0.780(2) & 0.756(2)  \\
5.93 & $20^3 40$ &  2.5 & 0.752(2) & 0.723(2)  \\
6.0  & $24^3 64$ &  2.5 & 0.715(2) & 0.683(2)  \\
     & $18^3 64$ &  1.9 & 0.717(7) & 0.679(7)  \\
     & $32^3 48$ &  3.1 & 0.713(1) & 0.679(2)  \\
6.2  & $32^3 64$ &  2.4 & 0.662(5) & 0.624(5)  \\
6.4  & $40^3 96$ &  2.3 & 0.643(7) & 0.607(7)  \\
     & $32^3 96$ &  1.8 & 0.658(12)& 0.613(11) \\
     & $48^3 96$ &  2.7 & 0.642(5) & 0.607(5)  \\
6.65 & $56^3 96$ &  2.3 & 0.635(7) & 0.606(7)  \\
\hline
\end{tabular}
\vspace*{-6mm}
\end{table}

\section{Run parameters and results}
Table~\ref{tab:run} lists the parameters of all our runs to date,
together with $B_K$(NDR, 2 GeV) obtained with gauge non-invariant
and invariant operators at each $\beta$.

Finite-size studies have been carried out at $\beta=6.0$ and 
6.4\cite{JLQCD2}. The 
finite size effect is less than 1\%
above 2.3~fm at both $\beta$, 
though it increases to 2\% below 2.0~fm at 
$\beta = 6.4$.
We have therefore focused our runs on a physical spatial lattice size of 
$La \approx 2.3-2.8$~fm. We expect the finite size effect to be less than 
our statistical errors.

\section{Continuum extrapolation}
\begin{figure}[bt]
\centerline{\epsfxsize=8.5cm \epsfbox{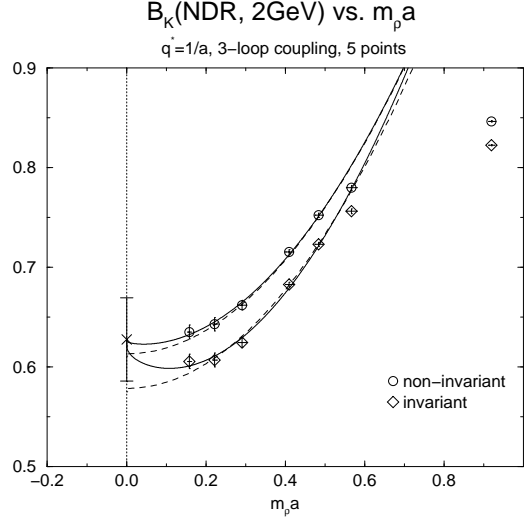}}
\vspace*{-12mm}
\caption{Gauge non-invariant(circles) and invariant(diamonds) 
$B_K$({\rm NDR}, 2GeV) as a function of 
$m_\rho a$, together with a simultaneous fit for the two operators 
including $\alpha^2$ term (solid lines)
and separate fits quadratic in $a$(dashed lines) 
to the five pairs of data points for $\beta\geq 5.93$.}
\label{bkmro}
\vspace*{-6mm}
\end{figure}

In Fig.~\ref{bkmro} we present $B_K$(NDR, 2 GeV) as a function of
$m_\rho a$.  The leftmost points are the new data taken at
$\beta=6.65$ on a $56^3\times 96$ lattice.
The new data points strongly support
an $O(a^2)$ behavior, the onset of which was noted in Lattice 96
at $\beta =$ 6.4.
The effect of higher order terms is not discernible for 
$\beta \ge 5.93$. 
Therefore, we fit the five points above $\beta=5.93$
with the form $B_K=c_0+c_1(m_\rho a)^2$,
which is shown by dashed lines in Fig.~\ref{bkmro}.
The continuum extrapolation gives
$B_K$(NDR, 2 GeV)=0.616(5) for the gauge
non-invariant operator, and 0.580(5) for the invariant one,
the average of the two being 0.598(5).

\section{Operator dependence}
\begin{figure}[bt]
\centerline{\epsfxsize=8.5cm \epsfbox{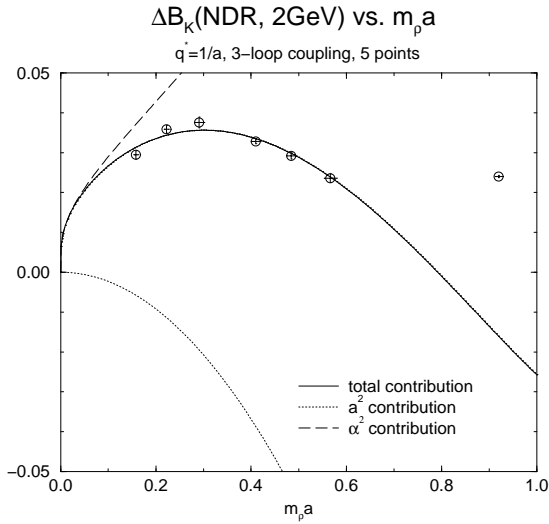}}
\vspace*{-12mm}
\caption{Difference of $B_K$({\rm NDR}, 2GeV) between
gauge non-invariant and invariant operators
as a function of $m_\rho a$.
The solid line represents a fit with $a^2$ and $\alpha^2$ terms,
while the dotted (dashed) line 
is the contribution from the $a^2$ ($\alpha^2$) term.}
\label{difBK}
\vspace*{-6mm}
\end{figure}

It has been noted that the two operators yield different values in the
continuum\cite{JLQCD1} (see Fig.~\ref{bkmro}).
After Lattice 97 a further analysis was made on this point, 
which we report in this write-up.
The difference between gauge non-invariant and invariant operators
should receive not only $O(a^2)$ scaling violation 
but also $\alpha_{\overline{MS}}(q^*)^2$ errors from the matching 
procedure.
Figure~\ref{difBK} plots this difference as a function of  $m_\rho a$.
Small errors resulting from a correlation between the two operators
allow us to fit the five data points with the form $b_1 (m_\rho a)^2 + b_2 
\alpha_{\overline{MS}}(q^*)^2$, giving $b_1=-0.23(2)$ and $b_2=1.73(5)$
with $\chi^2/{\rm d.o.f}=2.2$. The solid line indicates the fit, and others
show the breakdown into the $a^2$(dotted line) and $\alpha^2$(dashed line) 
contributions.
Allowing for a non-zero constant $b_0$, similar fitting
gives 
$b_0=-0.032(16)$, $b_1=-0.44(11)$ and $b_2=3.4(8)$.

Encouraged by this analysis 
we attempt to fit the five points at $\beta\ge 5.93$
simultaneously for both operators
including their correlations, 
with the form
$B_K^{\rm non-inv.}=c_0+c_1a^2+c_2\alpha_{\overline{MS}}(q^*)^2$
and 
$B_K^{\rm inv.}=d_0+d_1 a^2+d_2\alpha_{\overline{MS}}(q^*)^2$.
This fit yields $c_0=0.67(6)$ and $d_0=0.71(7)$, and hence we
impose the constraint  $c_0=d_0$ in our final fit.
In the continuum limit the fit (solid lines in Fig.~\ref{bkmro}) gives
$B_K$ (NDR, 2 GeV) = 0.628(42). The error is roughly ten times 
larger than the one from the quadratic fit. This originates from
uncertainties of the coefficient of $\alpha^2$ terms:
$c_2=-0.5(2.0)$ and $d_2=-2.2(2.0)$. 
The difference, however, is well constrained; $c_2 - d_2=1.7$ 
agrees well with $b_2=1.73$
obtained above.

We find larger coefficients $c_2=-1.0(4.2)$ and 
$d_2=-4.3(4.2)$ when $q^*=\pi/a$ is used,
or $c_2=1.6(1.5)$ and $d_2=-3.2(1.5)$
if mean-field improvement of the operators is not made. This
supports the argument of Ref.~\cite{LM}.

The final value depends only weakly on our choice of 
$\Lambda_{\overline{MS}}=0.23$~GeV: {\it e.g.,}
$B_K$(NDR, 2 GeV)=0.627(42) 
for  $\Lambda_{\overline{MS}}$=0.22~GeV and 0.628(41) for 0.24~GeV
in the continuum limit.

\section{Conclusions}
\vspace*{-1mm}
As our final value of $B_K$ in the continuum limit we adopt the 
result from the fit including the $\alpha^2$ term,
$
B_K ({\rm NDR, 2 GeV}) = 0.628 \pm 0.042,
$
which includes a systematic error due to the 2-loop uncertainty.
The size of the quoted error is 6.6\%, which roughly equals 
$3\times\alpha_{\overline{MS}}(q^*=1/a)^2$
at our smallest lattice spacing $1/a=4.87{\rm GeV}$ at $\beta =6.65$ 
where $\alpha_{\overline{MS}}(4.87{\rm GeV})=0.147$.
This magnitude of error 
is unavoidable 
unless a two-loop calculation is carried out
for the lattice renormalization.

Our $B_K$ is 
consistent with the JLQCD value obtained using the 
Wilson quark action,
$
B_K({\rm NDR, 2 GeV})=0.562\pm 0.064 ,
$
in which the operator mixing problem is solved
non-perturbatively with the aid of chiral Ward identities\cite{JLQCD3}.

This work is supported by the Supercomputer 
Project (No.~97-15) of High Energy Accelerator Research Organization (KEK),
and also in part by the Grants-in-Aid of 
the Ministry of Education (Nos. 08640349, 08640350, 08640404,
09246206, 09304029, 09740226).

\end{document}